\documentclass[aps, pra, singlecolumn]{revtex4-1}
\usepackage[T1]{fontenc}
\usepackage{lmodern}
\usepackage{amssymb}
\usepackage{amsmath}
\usepackage{siunitx}
\usepackage{graphicx,caption}
\usepackage[colorinlistoftodos,prependcaption,textsize=tiny]{todonotes}
\usepackage{hyperref}
\usepackage{soul}
\usepackage{blindtext}
\usepackage{setspace}
\presetkeys%
    {todonotes}%
    {inline,backgroundcolor=yellow}{}

\newcommand{\hzero}{H_0}    

\newcommand{\phiwall}{\varphi_{\rm wall}}
\begin{document}
\title{First principle simulation of ultra-cold ion crystals in a Penning trap with Doppler cooling and a rotating wall potential}
\author{Chen Tang$^1$, Dominic Meiser$^1$, John J. Bollinger$^2$ and Scott E. Parker$^1$}
\affiliation{ $^1$Department of Physics, University of Colorado at Boulder, Boulder, CO 80309 \\
$^2$National Institute of Standards and Technology, Boulder, Colorado 80305}

\begin{abstract}
  A direct numerical simulation of many interacting ions in a Penning trap with a rotating wall is presented.  The ion dynamics is modelled classically. Both axial and planar Doppler laser cooling are modeled using stochastic momentum impulses based on two-level atomic scattering rates. The plasmas being modeled are ultra-cold two-dimensional crystals made up of 100's of ions. We compare Doppler cooled results directly to a previous linear eigenmodes analysis. Agreement in both frequency and mode structure are obtained. Additionally, when Doppler laser cooling is applied, the laser cooled steady state plasma axial temperature agrees with the Doppler cooling limit. Numerical simulations using the approach described and benchmarked here will provide insights into the dynamics of large trapped-ion crystals, improving their performance as a platform for quantum simulation and sensing.
  
\end{abstract}

\maketitle  

\section{Introduction}
\label{sec:intro}

%{\bigskip \it SEP: Referring to the NIST Penning Trap requires references.  Take a look at Torrisi2016 for phrasing and Refs. 6,9,13 in that paper.}

A Penning trap with a rotating wall potential and Doppler laser cooling can produce stable ultra-cold non-neutral ion plasma crystals with temperatures of a fraction of a millikelvin \cite{jordan2018, tan1995, mavadia2013control, ball2018site}.  Such ultra-cold ion crystals enable interesting research at the forefront of different areas of physics including atomic physics \cite{gruber2001evidence, shiga2011diamagnetic, andelkovic2013laser,werth2018zeeman, stutter2018sideband}, quantum optics and metrology\cite{bohnet2016quantum, gilmore2017amplitude} , quantum simulation \cite{wang2013phonon, britton2012engineered, garttner2017measuring}, and basic plasma physics\cite{dubin1999trapped} , including studies of strongly coupled plasmas that model dense astrophysical matter\cite{van1991dense, ichimaru1987statistical, baiko2009coulomb} .  For many applications lower ion temperatures are desirable.  For example, lower ion temperatures can improve the fidelity of quantum simulations with trapped ions \cite{blatt2012quantum} and enable the engineering of stronger interactions by using shorter wavelength spin-dependent forces\cite{britton2012engineered} .  Colder temperatures also improve the stability of the ion crystal, which advances the prospects for single site optical manipulation and detection in multi-dimensional crystals, akin to what has been achieved with linear ion crystal arrays\cite{debnath2016demonstration} and with an ``atom microscope'' for neutral atoms\cite{bakr2009quantum}.  Single site optical manipulation and detection will enable the preparation of complicated entangled states through the implementation of variational quantum simulation protocols\cite{kokail2018self} on large trapped ion crystals.

The study of the thermodynamics of ultra-cold ions in Penning traps is not just a facilitator for certain experiments, it is also interesting basic plasma physics research in its own right. The ions in these systems form a strongly coupled plasma with complex collective modes of motion\cite{dubin1999trapped}. The behavior of energy transfer by nonlinear mode coupling, as well as other heating and cooling mechanisms is subtle and complex\cite{jensen2005rapid, dubin2005measurement, anderegg2009measurement}.

The primary means of cooling the ions in a Penning trap are various forms of laser cooling including Doppler cooling\cite{itano1982laser,itano1988perpendicular,torrisi2016perpendicular,asprusten2014theory}, side band cooling\cite{stutter2018sideband}, and electromagnetically induced transparency (EIT) cooling \cite{jordan2018, shankar2018modeling}. The basic principles for laser cooling ultra-cold ions are similar to that used for laser-cooling neutral atoms in traps, but there are important differences that can make it more challenging to fully understand the cooling limits and to optimize the laser cooling geometry and parameters such as laser intensity and detuning\cite{torrisi2016perpendicular}.  A significant difference from neutral atom experiments is that the ion motion is subject to the trapping electric and magnetic fields. In the axial magnetic field of the Penning trap, the ion plasma rotates, resulting in a change in the ion velocity relative to the cooling laser beam which is stationary in the laboratory frame. The typical velocity of the ions gives rise to Doppler shifts that can be large relative to the natural atomic line width of the ions. 

A second major difference is the strong Coulomb interaction between the ions. The Coulomb force couples the ions producing collective modes of motion. In contrast to the cooling dynamics of neutral atoms\textemdash which is largely a single particle phenomenon\textemdash it is necessary to take the collective nature of the ion motion into account to fully understand the cooling of the ion crystal. Even an analysis in terms of collective linear eigenmodes cannot fully account for the ion dynamics. Nonlinear coupling between modes gives rise to frequency shifts\cite{marquet2003phonon,mcaneny2014intrinsic} and broadening of resonances resulting in non-trivial frequency spectra.

In this paper we present results from a first principle classical model of the ion dynamics, analogous to a molecular dynamics simulation, including the trapping fields with a rotating wall potential, Coulomb interactions between ions, and Doppler laser cooling. We focus on the experimentally relevant case of a single-plane crystal generated with trapping parameters similar to that used in recent quantum simulation and sensing experiments\cite{bohnet2016quantum,gilmore2017amplitude,wang2013phonon,britton2012engineered,garttner2017measuring}.  These simulations provide a more complete picture of the Doppler cooling dynamics and cooling limits in the complex Penning trap geometry and can help with optimizing trap configuration and cooling parameters in current and future experiments. The simulation presented in this paper goes beyond prior modeling work\cite{wang2013phonon, itano1988perpendicular,torrisi2016perpendicular} in that it captures the combined effects of finite temperature, nonlinear coupling, and dynamics occurring at disparate time scales in a single unified simulation code.
The rest of this article is organized as follows. Section II discusses the simulation model, including the Penning trap Hamiltonian, Doppler laser cooling, the time integration of the ion equations of motion and numerical convergence with respect to timestep. We then compare the simulation results with both the axial and the planar linear eigenmode analysis in Sec. III. In Section IV, we show Doppler cooled simulations run to steady state and compare with results from equilibrium models\cite{itano1982laser,itano1988perpendicular,torrisi2016perpendicular}. We conclude with a summary and discuss future extensions of this work.

\section{Model and computational algorithm}
\label{sec:model}

In this section we describe our mathematical model and
computational approach for the simulation of ultra-cold ions in a Penning trap with rotating wall.  For a sketch of a Penning trap and a detailed discussion of the confining fields, see Ref. \cite{wang2013phonon}.

\subsection{Trap forces and Hamiltonian}

We treat the ions as classical point particles with velocity
$\mathbf{v}_i$ and position $\mathbf{x}_i$. Excluding the cooling
laser for now, the motion of the ions is governed by the
Hamiltonian
\begin{equation}
  H = \hzero + \sum_{i=1}^N q_i\varphi(\mathbf{x}_i,t)\;,
\end{equation}
where the Hamiltonian for the motion in the strong axial B-field is
\begin{equation}
  \hzero =
  \sum_{i=1}^N \frac{1}{2m_i}\left(
    \mathbf{p}_i -
    q_i\mathbf{A}(\mathbf{x}_i) \right)^2 \;,
  \label{eqn:total_hamiltonian}
\end{equation}
including the vector potential $\mathbf{A}$ for the homogeneous
axial magnetic field $\mathbf{B}=(0, 0, B_z)^T$ in the the
Penning trap. To be noted, in the following we use the terms axial, out-of-plane
and parallel as directions parallel to magnetic field, and planar, in-plane and perpendicular
as directions perpendicular to magnetic field. We choose our coordinate system such that the
Penning trap magnetic field $\mathbf{B} = \mathbf{\nabla}\times
\mathbf{A}$ is parallel to the $z$ axis. With that choice of
coordinate system we may choose $\mathbf{A} = -yB_z\mathbf{\hat x}$ with
$\mathbf{\hat x}=\mathbf{x}/x$ the unit vector along
$\mathbf{x}$. In Eq.~\eqref{eqn:total_hamiltonian}, $N$ is the
number of ions in the trap and $m_i$ and $q_i$ are the mass and
charge of ion $i$. The electrostatic potential
\begin{equation}
  \varphi(\mathbf{x}_i, t) =
  \varphi_{\rm{trap}} (\mathbf{x}_i) +
  \phiwall(\mathbf{x}_i, t) +
  \sum_{j \neq i}
  \frac{1}{8\pi\varepsilon_0}\
    \frac{q_j}{\left| \mathbf{x}_i - \mathbf{x}_j \right|}
\end{equation}
contains the potential $\varphi_{\rm{trap}}$ for the electrodes in the
Penning trap, the rotating wall potential $\phiwall$, and the
Coulomb potential for the interaction between the ions. In the
vicinity of the ion crystal the trap potentials are well approximated
by harmonic potentials. We parametrize $\varphi_{\rm{trap}} (\mathbf{x})$ as
\begin{equation}
  \varphi_{\rm{trap}}(\mathbf{x})  =\frac{1}{4}k_z \left(2z^2 - x^2 - y^2\right),
  \label{equ:original_lab}
\end{equation}
and the time-dependent rotating wall potential $\phiwall(\mathbf{x}, t)$ as
\begin{equation}
  \phiwall(\mathbf{x}, t) =\frac{1}{2}k_z\delta\left( x^2 + y^2\right)\cos\left [2\left(\theta +\omega_R t\right)\right] \, ,
  \label{equ:rotating_lab}
\end{equation}
where $\delta$ is the dimensionless parameter that characterizes the strength of the rotating wall potential to the trapping potential $\varphi_{\rm{trap}} (\mathbf{x})$ and $\theta$ is the azimuthal angle in cylindrical coordinates
.
Next we transform to the rotating frame
\begin{equation}
\left[
\begin{array}{c}
x_R\\
y_R
\end{array}\right] =
\left[
\begin{array}{cc}
\cos(\omega_R t) & -\sin(\omega_R t)\\
\sin(\omega_R t) & \cos(\omega_R t)
\end{array}\right]
\left[\begin{array}{c}
x\\
y
\end{array}\right]\;.
\label{equ:transfer}
\end{equation}
 Using the above transformation, Eq.~(\ref{equ:transfer}), to the rotating frame, and combining Eqs.~(\ref{equ:original_lab})~and~(\ref{equ:rotating_lab}) we obtain
\begin{equation}
  \varphi_{\rm trap,R}(\mathbf{x}_R) +\varphi_{\rm wall, R}(\mathbf{x}_R) =\frac{1}{2}k_z z^2 - 
\frac{1}{2} \left(k_x x_R^2 + k_y y_R^2\right) \, 
\end{equation}
where
\begin{equation}
k_x=k_z(\frac{1}{2}-\delta),\qquad 
k_y=k_z(\frac{1}{2}+\delta)\; .
\end{equation}
Note that in the rotating frame, the applied potential is time independent, Eq.~(7). Due to the rotating wall potential,
Eq. (5), the externally applied trapping fields produce a Hamiltonian that is time dependent in the laboratory frame.  However, the applied potential is time independent in the rotating frame (Eq. 7), resulting in energy conservation in this
frame\cite{dubin1999trapped}. This is a useful check for the validity of the model in the absence of laser cooling. The trap Hamiltonian in the rotating frame, $H_R$, is
\begin{equation}
\begin{aligned}
 H_R =
&\sum_{i=1}^N \frac{1}{2m_i}\left(
\mathbf{p}_{Ri} -
q_i\mathbf{A}(\mathbf{x}_{Ri}) \right)^2 \, +\sum_{i=1}^N q_i\left(\varphi_R(\mathbf{x}_{Ri}) +
\phi_{\rm wall, R}(\mathbf{x}_{Ri})\right) \, +\\
&\sum_{i=1}^N \frac{1}{2}\left(q_i B_z \omega_R-m_i \omega_{R}^{2}\right) r_{Ri}^{2}  \, + 
\sum_{i=1}^N q_i \sum_{j \ne i}\frac{1}{8\pi\varepsilon_0}\frac{q_j}{\left| \mathbf{x}_{Ri} - \mathbf{x}_{Rj} \right|}
\end{aligned}
\end{equation}
where all ''R'' subscripts indicate the rotating frame and $r^2_{Ri} = x^2_{Ri} + y^2_{Ri}$. The third term is new and is due to the Lorentz and centrifugal forces. The Coulomb interaction energy term only involves the difference between the coordinates of the ions.  The transformation to the rotating frame conserves the length of vectors, so the expression is the same in the rotating or lab frame\cite{dubin1999trapped}. After the system reaches equilibrium, all the particles nearly rigidly rotate as a crystal with the constant frequency $\omega_R$. In Sec.~IV., the measurement of temperature via kinetic energy is made in the rotating frame so as not to include the energy associated with the rigid rotation of the ion crystal.

%Since, all ions rotate together uniformly, the Coulomb interaction, last term in Eq.~(3), is unchanged, and hence energy is conserved for the system. Throughout this paper, all measurements are made in the rotating frame as defined by Eq.~(6).

\subsection{Doppler cooling}

In addition to the conservative dynamics described by the
Hamiltonian, the ions are subject to radiation pressure forces
due to cooling lasers. Our model of laser cooling is based on
resonance fluorescence for a driven two-level atom. We begin by
discussing our approach for a single cooling laser.  The photon
scattering rate for a driven two-level atom is given by\cite{itano1988perpendicular}
\begin{equation}
\dot n (\mathbf{x}, \mathbf{v}) = 
S(\mathbf{x})\gamma_0
\frac{(\gamma_0/2)^2}{(\gamma_0/2)^2(1+2S(\mathbf{x}))+\Delta^2(\mathbf{v})}\;,
\end{equation}
where $\gamma_0$ is the natural line-width of the atom transition
(in radians per second), $S(\mathbf{x})$ is the saturation
parameter, and $\Delta(\mathbf{v})=\Delta_0 -
\mathbf{k}\cdot\mathbf{v}$ is the detuning of the atomic
transition from the laser frequency including the first order
Doppler shift with $\Delta_0$ the detuning at rest. The vector
$\mathbf{k}$ is the wave vector of the cooling laser. We assume
that the atoms scatter photons with this rate with Poissonian
number statistics. The saturation parameter is spatially
dependent due to the intensity variations of the cooling laser
beam. For a Gaussian beam we have
\begin{equation}
S(\mathbf{x})=S_0e^{-\rho^2/W_y^2}\;,
\end{equation}
where $\rho$ is the distance of the atom from the axis of the
beam and $W_y$ is the $1/e$ radius of the intensity of the beam.

To incorporate laser cooling into our numerical simulations we
proceed as follows. First, we compute the mean number of photons
scattered by ion $i$ in time interval $\Delta t$,
\begin{equation}
\bar{n}_i=\dot{n}_i \Delta t\;.
\end{equation}
The velocities and positions needed for computing $\dot{n}_i$ are
evaluated at the center of the time step in accordance with the
integration scheme discussed below in Sec.~II.C. We then compute the actual
number of photons scattered by each ion, $n_i$, as a Poisson
random number with mean $\bar{n}_i$. Each ion receives a total
momentum kick of 
\begin{equation}
\Delta \mathbf{p}_i^{\rm Laser} = \Delta \mathbf{p}_{i,{\rm absorb}} + 
\Delta \mathbf{p}_{i,{\rm emit}}\;,
    \label{eqn:dopplerkick}
\end{equation}
where $\Delta \mathbf{p}_{i,{\rm absorb}}=n_i \hbar \mathbf{k}$
and $\Delta \mathbf{p}_{i,{\rm emit}}$ is the recoil
corresponding to $n_i$ photons scattered in random directions
with an isotropic probability distribution. To compute
$\Delta\mathbf{ p}_{i,{\rm emit}}$, we generate $n_i$ vectors of
length $\hbar k$ pointing in random directions.  The recoil
momentum $\Delta\mathbf{ p}_{i,{\rm emit}}$ is then obtained by
adding up these vectors.

For simulating multiple cooling lasers we find the momentum
kick~Eq.~(\ref{eqn:dopplerkick}) for each laser individually
and then we add up the results. This approach captures the salient features of laser cooling with
the following approximations. First, in the case of
strong saturation, $S\gtrsim 1$, the momentum kicks from multiple
beams are not additive but are correlated in a more complicated
way. We neglect these correlations. The other approximation in
our model is that we assume that the ion motion is uniform during
an excited state lifetime. For motion of ions in the Penning trap
the most rapid change of the ion velocity is due to the magnetic
field. We can characterize the validity of the assumption of
uniformity of motion by the dimensionless parameter
$\eta=\omega_B/\gamma_0$ being small compared to $1$. The
parameter $\omega_B=B_zq/m$ is the Larmor precession frequency.
For the trap and ion parameters considered in this paper we have
$\eta \sim 0.3$. We also assume that the cooling laser intensity
is approximately constant during an excited state lifetime. For
the typical parameters considered in this publication this
approximation holds to a higher degree of accuracy than the
assumption of a constant ion velocity during an excited state
lifetime.

\subsection{Time integration of the equations of motion}

%{\bigskip \it 

%\bigskip
%Buneman67 - O. Buneman, J. Comput. Phys. {\bf 1} 517 (1967).

%\bigskip
%Boris70 - J.P. Boris, Proceedings of 4th Conference on Numerical Simulation of Plasmas, Naval Research Laboratory, Washington D.C., 1970, pp. 3-67.

%\bigskip
%Birdsall85 - C.K. Birdsall and A.B. Langdon, ``Plasma physics via computer simulation,'' p. 62, McGraw-Hill (1985).

%\bigskip}

The numerical integration of the ion motion is performed in the lab frame of reference. A time splitting technique is used to numerically integrate the equations of motion for the ions.
This is analogous to the Buneman and Boris algorithms \cite{buneman1967time,boris1972proceedings,birdsallplasma} with an exact matrix rotation for the $\mathbf{v} \times \mathbf{B}$ force term. To advance the positions and
velocities of the ions, $\{\mathbf{x}_i, \mathbf{v}_i\}$, from
time $t$ to $t + \Delta t$ we apply the following time evolution operator
\begin{equation}
    \{ \mathbf{x}_i(t+\Delta t), \mathbf{v}_i(t+\Delta t) \} =
    U(\Delta t)
    \{\mathbf{x}_i(t), \mathbf{v}_i(t)\}\;,
\end{equation}
where $U(\Delta t)$ is given by
\begin{equation}
  U(\Delta t) = 
  U_{0}(\Delta t /2)
  U_{\rm kick}(t + \Delta t / 2; \Delta t)
  U_{0}(\Delta t /2)\;.
\end{equation}
 $U_{0}(\Delta t/2)$ is the time evolution
operator corresponding to $\hzero$  (see Eq. (2)) that advances the state of the
ions for a time interval of duration $\Delta t / 2$. Since $\hzero$
 contains just the kinetic energy of the ions and
the Lorentz force due to the axial magnetic field. the motion
generated by $U_{0}$ is given by Larmor
precession,
\begin{eqnarray}
  &\mathbf{x}_i(t + \Delta t / 2) =
                                   \mathbf{x}_i(t)+
      \left[\begin{array}{ccc}
        s & c - 1 & 0\\
        -(c - 1) & s & 0\\
        0 & 0 & \omega_B\Delta t / 2
      \end{array}\right]\frac{\mathbf{v}_i(t)}{\omega_B}\;,\nonumber\\
 & \mathbf{v}_i(t+\Delta t/2) = \left[\begin{array}{ccc}
        c & -s & 0\\
        s & c & 0\\
        0 & 0 & 0
      \end{array}\right]\mathbf{v}_i(t)\;,
\end{eqnarray}
where
\begin{equation}
    s = \sin(\omega_B \Delta t / 2)\quad \hbox{and} \quad
    c = \cos(\omega_B \Delta t / 2)\;.
\end{equation}

The time evolution operator $U_{\rm kick}(t + \Delta t/2;
\Delta t)$ captures the forces due to
the electrostatic potential (see Eq. (3)) as well as the radiation pressure
forces due to laser cooling (Sec. II.B). This operator is time
dependent. We evaluate it at the mid-point of the time interval,
$t + \Delta t / 2$. The operator $U_{\rm kick}$ changes
the velocities of the particles only,
\begin{eqnarray}
    \mathbf{v}_i(t + \Delta t) &=&
      \mathbf{v}_i(t) +
      \Delta t (q_i/m_i) \mathbf{E}{(t + \Delta t / 2, \mathbf{x}_i(t + \Delta t / 2))} +
      \Delta \mathbf{p}_i^{\rm Laser}/m_i
      \;,
      \nonumber
\end{eqnarray}
where
\[
    \mathbf{E}(t+\Delta t /2,\mathbf{x}_i(t + \Delta t / 2)) =
\mathbf{\nabla}\varphi(\mathbf{x}_i(t + \Delta t / 2))\;.
\]

\subsection{Convergence with respect to timestep}

\begin{figure}
  \includegraphics{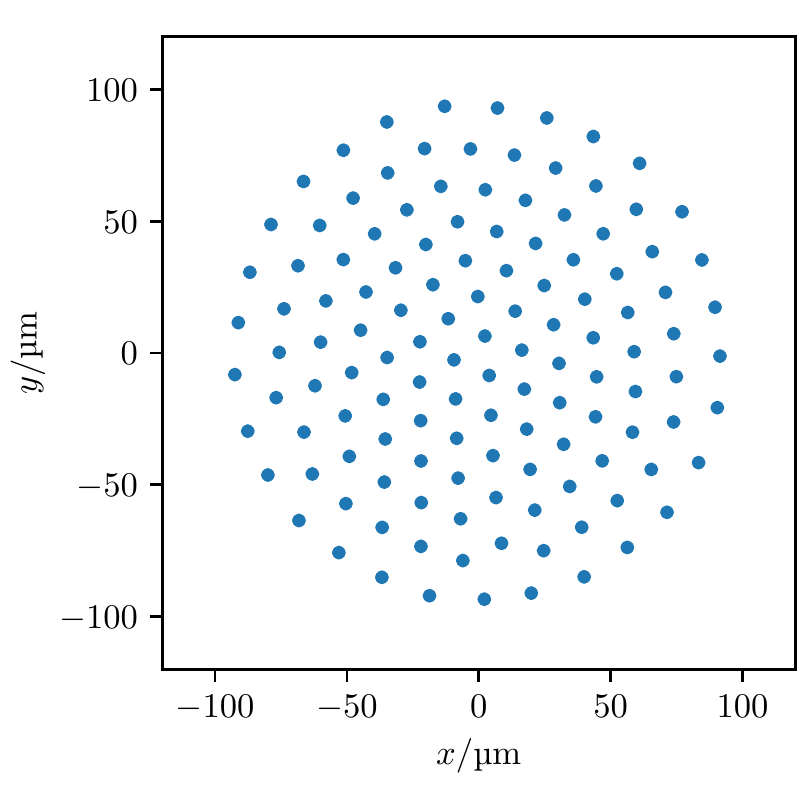}
  \caption{Top view of the steady state configuration of 127 ions in
    a Penning trap used for the convergence study.}
  \label{fig:initial_state_top_view}
\end{figure}

To verify the precision of our numerical integration procedure,
we check that the simulations converge with the expected
quadratic rate as the timestep size is reduced. To evaluate the
convergence we initialize our simulation with an equilibrium 
configuration for 127 ions shown in
Fig.~\ref{fig:initial_state_top_view}. This is the lowest energy state equilibrium (zero temperature 2D crystal) in the rotating frame
generated using the procedure previously discussed by Wang, et al. \cite{wang2013phonon}. The simulation is in the laboratory frame and so tracks the 
full dynamics, including the cyclotron and magnetron motion and so maintaining the 2D crystal equilibrium is a good test of the numerics. We study numerical convergence without Doppler laser cooling.
Here and in the rest of
the paper, unless indicated otherwise, we use typical parameters
for the Penning trap at the National Institute of Standards and Technology (NIST)\cite {bohnet2016quantum,britton2012engineered,sawyer2014spin} with a strong homogeneous magnetic
field of $B_z=\SI{4.4588}{\tesla}$, a trap rotation frequency of
$\omega_{\rm trap}=2\pi\times \SI{180}{\kilo \hertz}$, end cap
voltages yielding a confining potential of
$k_z=\SI{9.21}{\mega\volt/\meter^2}$, and a $\SI{1}{\volt}$ rotating wall
potential yielding
$\delta=\SI{3.5e-4}{}$. For $^9$Be$^+$ ions, which is assumed throughout this paper,
the cyclotron frequency is $qB/m = 2\pi\times \SI{7.596}{\mega\hertz}$ and the axial confining frequency parallel to the magnetic field is $2\pi\times \SI{1.58}{\mega\hertz}$.

Starting from this initial 2D crystal configuration we
integrate the equations of motion for a total duration, $\tau$,
using different time step sizes, $\Delta t$, to find the final
positions $\mathbf{x}_i(\tau; \Delta t)$. For each time step size we
compare the final solution with a reference solution,
$\mathbf{x}_i^{\rm ref}(\tau)=\mathbf{x}_i(\tau, \SI{2e-10}{\second})$,
computed using a time step size of $\Delta t =
\SI{2e-10}{\second}$. We compute the error $\Delta x(\tau; \Delta
t)$ as the average Euclidean distance of the positions from the
reference solution,
\begin{equation}
  \Delta x(\tau;\Delta t) =
  \sqrt{N^{-1}\sum_{i=1}^N\left(
      \mathbf{x}_i(\tau; \Delta t)-\mathbf{x}_i^{\rm ref}(\tau)\right)^2}\;.
\end{equation}

\begin{figure}
  \includegraphics{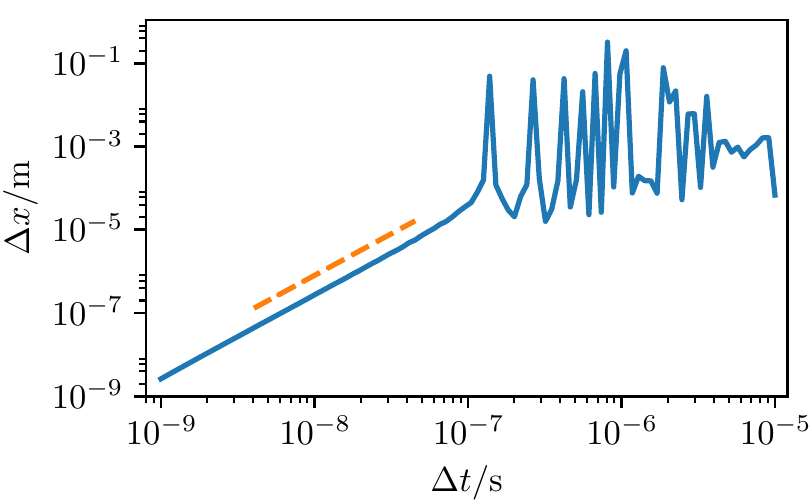}
  \caption{Time integration errors as a function of time step
    size for a total integration time of $\tau=\SI{10}{\us}$
    for a typical Penning trap simulation. The dashed orange line
    is a quadratic for orientation. See text for details.}
  \label{fig:convergence}
\end{figure}

Figure~\ref{fig:convergence} shows the integration errors for an
integration time of $\tau=\SI{10}{\us}$. The error
decreases quadratically for time step sizes below approximately
$\SI{1.0e-7}{\second}$. At a time step size of $\Delta t =
\SI{1}{\nano\second}$, the mean position error of the ions is
$\Delta x(\SI{1}{\nano\second})\approx \SI{1}{\nano\meter}$.
During this time interval the ions move on the order of
$\SI{1}{\milli\meter}$, i.e. the ion motion is integrated with a
relative error on the order of $\SI{1.0e-6}{}$. For time step
sizes greater than $\Delta t\approx \SI{1.0e-7}{\second}$
resonances occur where the integrator step size matches the
period of one of the vibrational eigenmodes of the system. At
these resonances the time integration errors can grow large.
Results presented in this paper were obtained with $\Delta t
= \SI{1}{\nano\second}$.

\section{Comparison with linear theory near zero temperature}
\label{sec:PSD}

We now turn to an analysis of the vibrational eigenmodes of the
2D ion crystal. To find the eigenmodes we initialize the ions in a
low energy spatial configuration found by minimizing the
potential energy in the rotating frame. We subject the ions to cooling lasers with a geometry
typical for the NIST experiment\cite{bohnet2016quantum,britton2012engineered,sawyer2014spin,torrisi2016perpendicular}. Details of the setup of the cooling
lasers are described in section~\ref{sec:Doppler}. The cooling
time for both in-plane and out-of-plane degrees of freedom is on
the order of $\SI{1}{\milli\second}$. After
$\SI{5}{\milli\second}$ of laser cooling the ions have settled
into an equilibrium state. We then turn off the cooling lasers
and record the trajectories of the ions $\mathbf{x}_i(t_j)$ at
discrete times $t_j = j\Delta t$ for $j=0,1,\ldots,N_{\rm
sample}-1$.

From the trajectories we compute power spectra of the ion motion
by means of
\begin{equation}
  P_z(\omega) =
    \sum_{i=1}^N\left| \tilde{z}_i(\omega) \right|^2 +
    \sum_{i=1}^N\left| \tilde{z}_i(-\omega) \right|^2\;.
    \label{eqn:psd}
\end{equation}
In this equation, $\tilde{z}_i$ is the Fourier transform of the
$z$ coordinate of ion $i$,
\begin{equation}
  \tilde{z}_i(\omega) = \frac{1}{\tau}\int_0^\tau e^{-i\omega
    t}z_i(t)\, \hbox{d}t,
\end{equation}
with $\tau = N_{\rm sample}\cdot\Delta t$ the total integration time
of the simulation.  In terms of the discretely sampled
trajectories, the Fourier transform can be approximated by a
discrete Fourier transform, which we evaluate with the aid of a
Fast Fourier Transform numerically,
\begin{eqnarray}
    \tilde{z}(l\Delta\omega)&=&
    \frac{1}{\tau}\int_0^\tau e^{-il\Delta\omega t}z(t)\,\hbox{d}t \\
    &\approx &\frac{1}{N_s}\sum_n^{N_s}e^{-i2\pi ln/N_s}z(n \Delta t)\;.
\end{eqnarray}
Here we have introduced the frequency resolution
$\Delta\omega=2\pi/\tau$ and $l=0,1,\ldots,N_{\rm sample} - 1$.
Frequencies exceeding the Nyquist frequency,
$l\Delta\omega>\pi/\Delta t$, wrap to negative frequencies. 

We compute spectra $P_x$ and $P_y$ for the in-plane degrees of
freedom similarly with a minor caveat: to eliminate the large
coherent component due to the uniform rotation in the trap we use
the coordinates in the rotating frame, $x_R$ and $y_R$.

\subsection{Out-of-plane eigenmodes}

\begin{figure}
    \includegraphics[width=8cm]{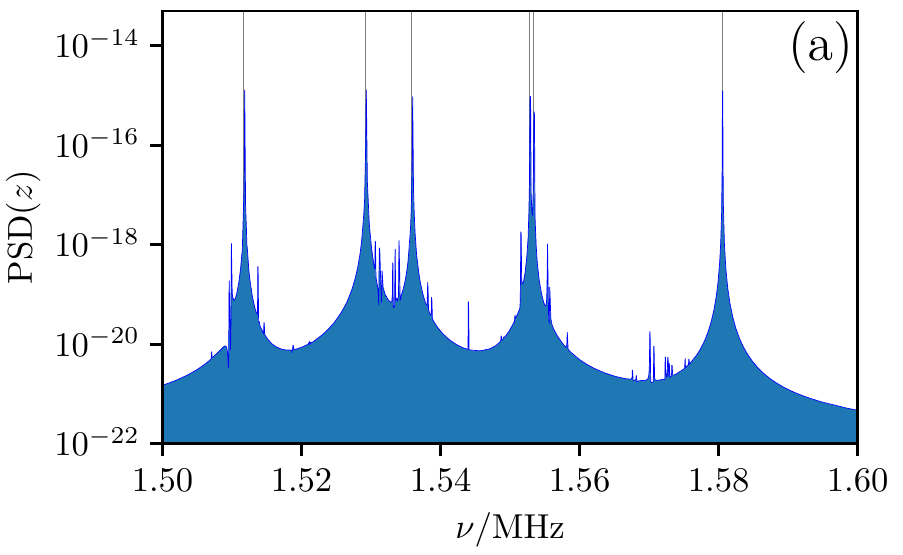}
    \includegraphics[width=8cm]{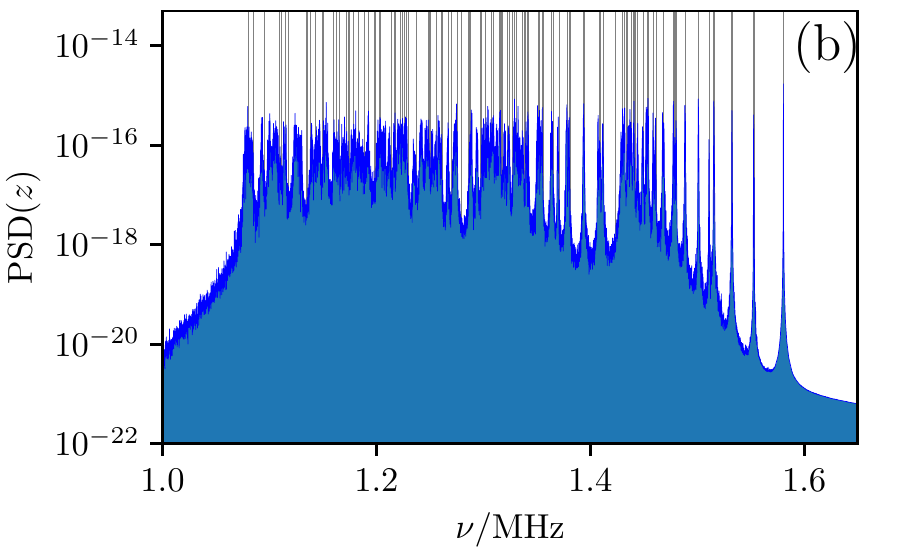}
    \caption{Spectra $P_z(\omega)$ of out-of-plane motion for 7
    ions (a) and 127 ions (b). Some modes are nearly degenerate and individual peaks are not discernible.
    \label{fig:Axial7}
    \label{fig:Axial127}
    }
\end{figure}

Figure~\ref{fig:Axial7}(a) shows $P_z$ for 7 ions with a
sampling period of $\Delta t = \SI{0.25}{\micro\second}$ and a
total integration time of $\tau=\SI{50}{\milli\second}$
corresponding to a frequency resolution of
$\Delta\omega/(2\pi)=\SI{20}{\hertz}$. We superimpose the mode
frequencies found by linearizing the equations of motion around
the ground state as grey vertical lines. For the linear analysis (grey lines) 
we use a code developed by Wang {\it et
al.}~\cite{wang2013phonon} with minor modifications. We observed that the
resonance frequencies obtained with our molecular dynamics
simulations agree well with the modes from the linear theory.
Several pairs of modes are nearly degenerate. For example, the
degeneracy of the two tilt modes around $\SI{1.55}{\mega\hertz}$ is
lifted only by the weak perturbation of the rotating wall
potential.  The out-of-plane eigenmodes are frequently referred to as drumhead modes because they resemble the vibrations of a membrane with open boundary conditions on the edge of the membrane.

Figure~\ref{fig:Axial127}(b) shows $P_z$ for 127 ions with
otherwise identical parameters. Again, the frequencies of the
highest frequency modes agree well with the zero temperature linear 
mode analysis. However, the frequencies of several of the lower frequency,
shorter wavelength modes deviate from the linear frequencies. The
reason for these shifts is that, at finite temperature, defects
form in the ion crystal, especially for larger numbers of ions. Due to their shorter wavelength, the lower frequency modes more sensitively depend on the local crystal structure, and therefore are perturbed by the crystal defects. In Sec.~IV.,
 we will discuss how the power spectrum in Fig.~\ref{fig:Axial127}(b) can be used to estimate the temperature of the drumhead modes.
%And the lower frequency modes, which depend more on local structure due to their shorter wavelength, are more sensitive to perturbations due to these crystal defects.

\begin{figure}
\includegraphics[width=8cm]{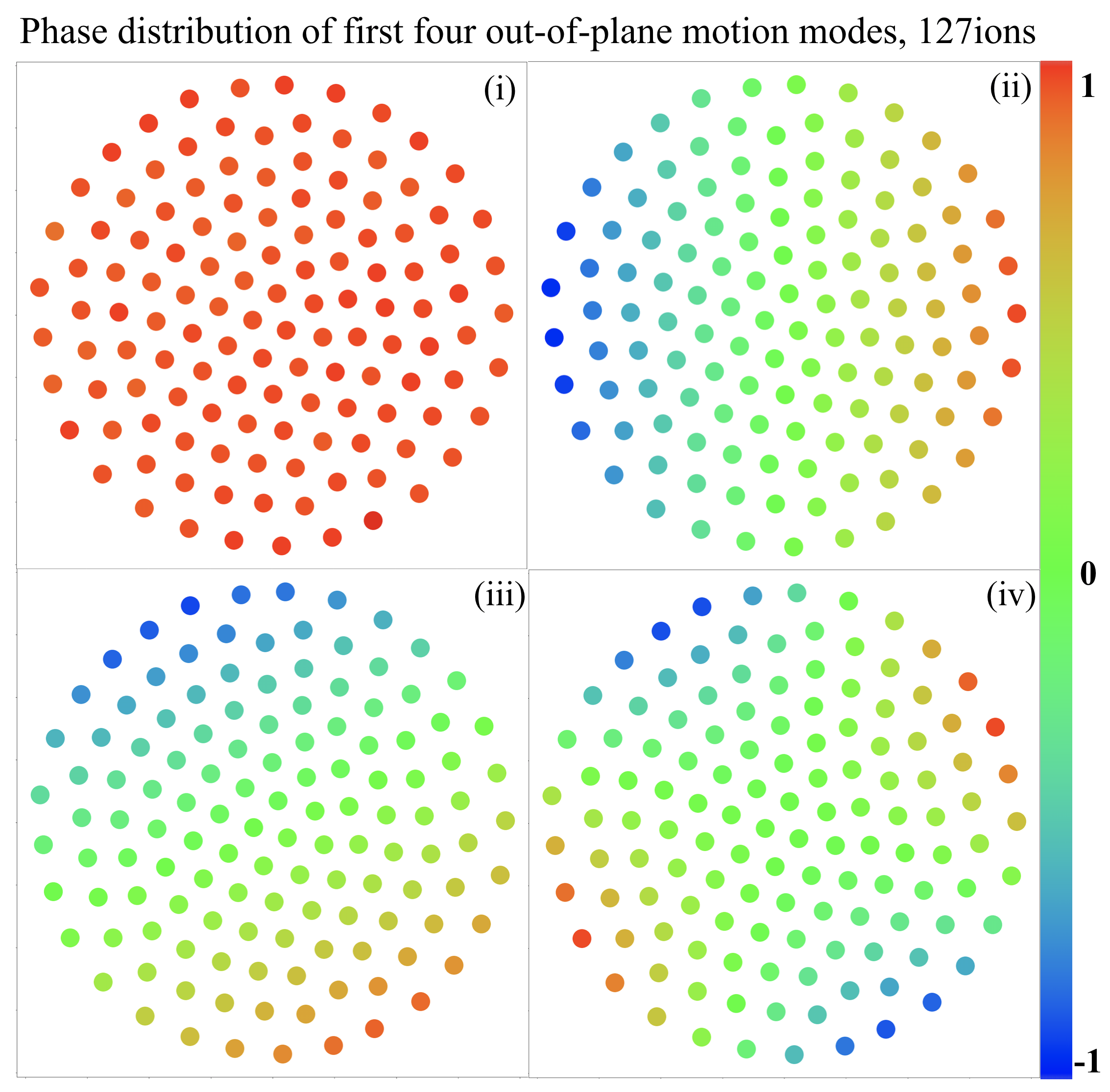}
 \caption{Vibrational mode patterns (eigenfunctions) for the first 4 out-of-plane modes using 127 ions.  Each dot represents the ion position, $z_i$.  The color represents the scale value of the normalized out-of-plane motion of the particular ion, as described in Sec.~III.A.}
 \label{fig:phase_axial127}
\end{figure}

In addition to frequency spectra, the axial vibrational mode patterns can be analyzed\cite{wang2013phonon}.  We obtain the eigenfunction shapes using a notch filter of width $\Delta \omega$ about the mode frequency of interest.  We then inverse Fourier transform $\tilde{z}_i(\omega)$ and plot the normalized displacement as a function of time.
Figure~\ref{fig:phase_axial127} shows a snaphot of the first four (starting from highest frequency) vibrational modes for 127 ions. 
Figure~\ref{fig:phase_axial127} (i) shows the center-of-mass mode where all ions move together in the $z$-direction. Figures~\ref{fig:phase_axial127} (ii) and (iii) show the next two highest frequency modes which are tilt modes that would be degenerate except for the weak structural anisotropy produced by the rotating wall potential. The fourth mode, showed in Fig.~\ref{fig:phase_axial127}(iv), has two linear nodes.  Figure~\ref{fig:phase_axial127} shows similar results as obtained in the eigenfunction plots presented in Ref.~\cite{wang2013phonon}, Fig.11.

\subsection{In-plane spectrum}

The spectrum of in-plane modes can be decomposed into two
qualitatively different types of modes. There is a high frequency
branch of modes near the cyclotron frequency and a low frequency
branch near zero frequency. The high frequency modes are
commonly referred to as cyclotron modes and the low frequency
modes are referred to as $E\times B$ modes. The spectra $P_x$ and
$P_y$ are very similar. Therefore we consider just $P_x$ for
simplicity.

%{\bigskip \it
%CT: (Do we need to add this line?) 7 and 127 ions cases are studied again, but for overview figure we only present
%7 ions in-plane PSD, Since modes are dense in overview plot and single branch display details of mode distribution better.
%\bigskip }

\begin{figure}
\includegraphics[width=8cm]{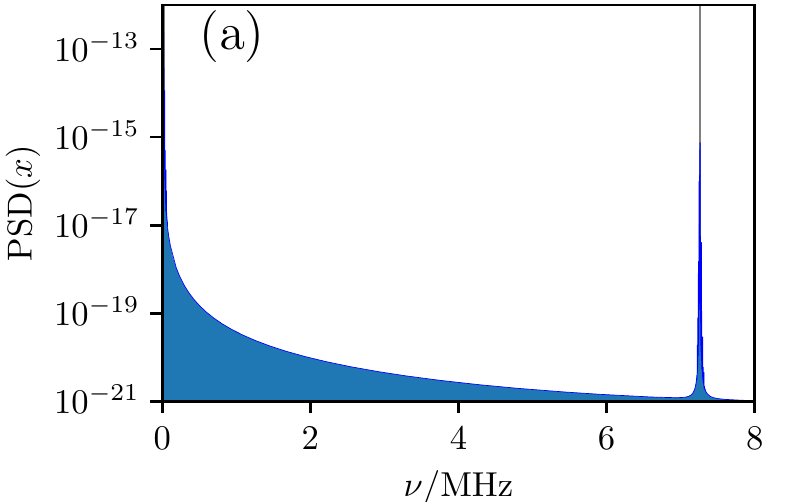}
 \includegraphics[width=8cm]{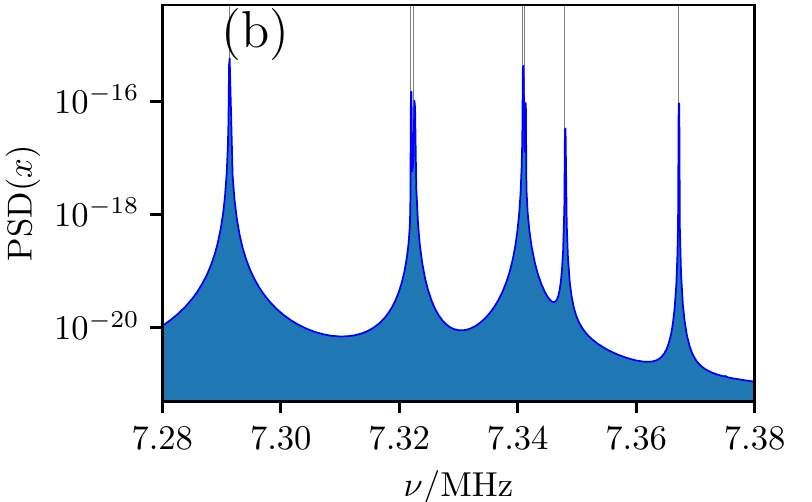}
 \includegraphics[width=8cm]{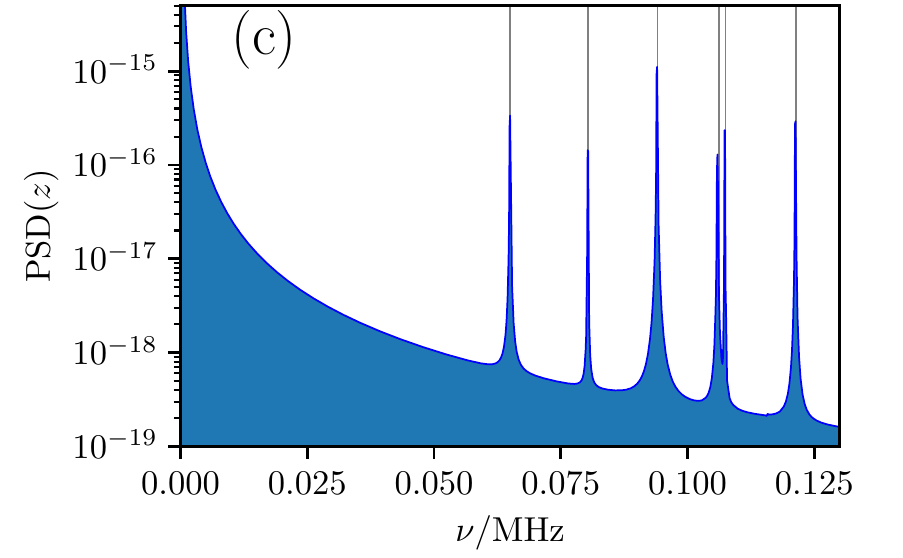}
 \caption{Spectra of the in-plane motion of a 7 ion crystal, $P_x(\omega)$. The top figure (a)
    shows the entire frequency range with the $E\times B$ modes near
    zero frequency and the cyclotron modes near $\omega_{B}=eB/m = (2\pi)\times 7.596$ MHz.
    The two figures $(b)$ and $(c)$ show closeups of the cyclotron (upper branch) and
    $E\times B$ (lower branch) modes, respectively. For this 7 ion case, we use a different trap axial frequency, to better display the cyclotrons modes. 
 \label{fig:7planar}
 \label{fig:7planar_upper}
 \label{fig:7planar_lower}
    }
\end{figure}

\begin{figure}
 \includegraphics[width=8cm]{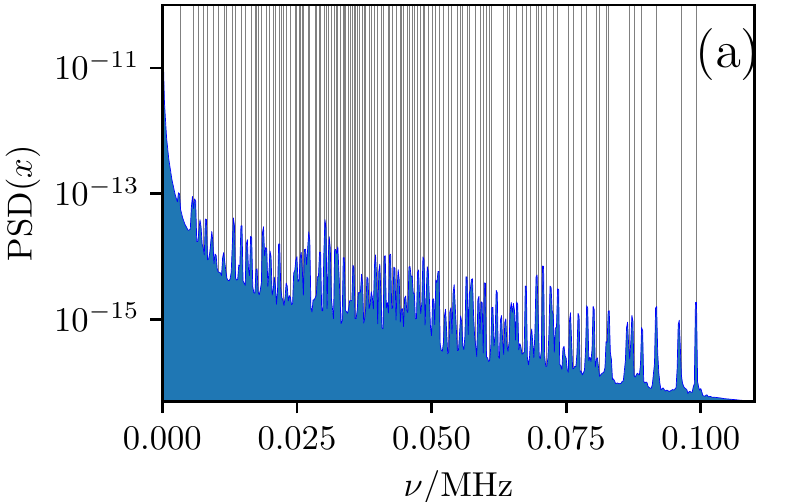}
 \includegraphics[width=8cm]{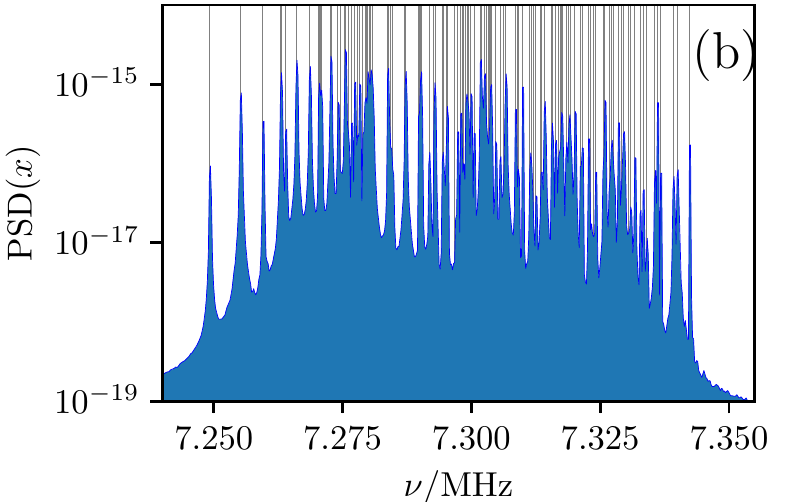}
 \caption{Two branches of the in-plane motion spectrum for a 127 ion crystal. Figures $(a)$ and $(b)$ show closeups of the  $E\times B$(lower branch) and
    cyclotron(upper branch) modes, respectively.
 \label{fig:127planar_lower}
 \label{fig:127planar_upper}
    }
\end{figure}

Figure~\ref{fig:7planar}(a) shows the
two narrow bands of modes for 7 ions together on the same plot. The mode frequencies
predicted by the linear theory are again superimposed as grey
vertical lines. Figures~\ref{fig:7planar_upper} (b) and
~\ref{fig:7planar_lower} (c) show the cyclotron and
$E\times B$ modes separately in more detail.  The left most peak near zero frequency shown in Fig.~\ref{fig:7planar_lower}(c), which is relatively higher than the other peaks, is associated with the slow periodic deformation of the elliptical ion crystal due to the confining potential of the rotating wall. This mode, sometimes called a rocking mode, is a zero frequency mode in the absence of a rotating wall. The presence of a wall potential leads to a small non-zero frequency for this mode.
Figures~\ref{fig:127planar_upper}(a) and ~\ref{fig:127planar_lower}(b) show the two branches of in-plane modes for the 127 ion crystal. %This matching behavior is similar to what we observed in Figure~\ref{fig:Axial127}(b) for sharing the same reason.%
Again, we observe good agreement between the zero-temperature linear analysis and the 
finite temperature direct numerical simulations. 

\section{Doppler cooling geometry and simulation results}
\label{sec:Doppler}

We now turn to an analysis of Doppler cooling of the ion motion. The numerical implementation of Doppler laser cooling was discussed in Sec.~II.B. Doppler laser cooling with a rotating wall is discussed in detail in Ref. \cite{torrisi2016perpendicular}.  There, it was shown in a fluid model that assumed thermal equilibrium that low ion temperatures close to the Doppler laser cooling limit was possible over a range of experimental conditions.  This is in contrast with a Doppler laser cooling model without a rotating wall \cite{itano1988perpendicular}, where the equilibrium temperature depended sensitively on the details of the set-up.
%Laser cooling in the planar direction involves balancing of the torque from the laser beam exactly with the torque imparted by the rotating wall\cite{torrisi2016perpendicular} to minimize heating from the perpendicular laser.  In the absence of the rotating wall, the applied torque balances with the  torque of the trap error fields resulting in temperatures that are well above the Doppler cooling limit\cite{itano1988perpendicular}.%
Here, we discuss in more detail, the geometry of the laser beams and the achievement of a Doppler cooled ultra-cold steady-state 2D crystal of 127 ions using direct numerical simulation.
Measurements of the average temperatures of the ion motion in the planar (perpendicular to the B-field) and axial (parallel to the B-field) directions are presented. In addition, mode resolved temperatures in the axial direction is investigated.

\subsection{Doppler cooling in a rotating wall Penning trap}

We consider a cooling laser configuration similar to that employed in a typical
NIST Penning trap experimental setup\cite{bohnet2016quantum,britton2012engineered,sawyer2014spin,torrisi2016perpendicular}. We assume the out-of-plane motion is cooled by two counter-propagating cooling beams along the positive
and negative $z$ directions, i.e.  parallel and anti-parallel to
the trap magnetic field. These two beams have equal intensity and
they are detuned by $\Delta_{\parallel}=-\gamma_0/2$ relative to the atomic
transition. The waist of the beams is assumed to be
larger than the size of the ion crystal so that the 
spatial variation of the laser intensity can be neglected.

\begin{figure}
  \includegraphics[width=8cm]{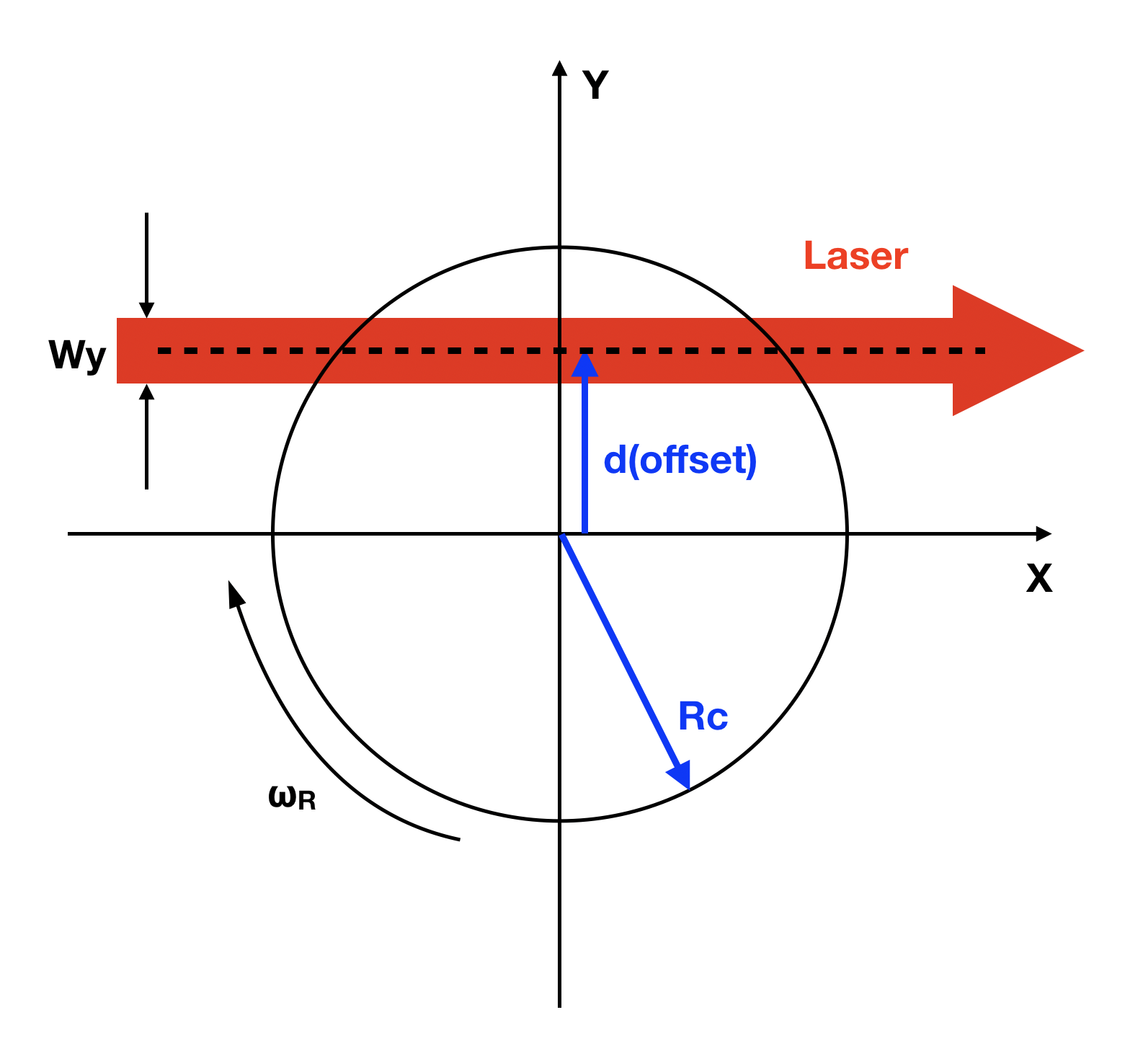}
  \caption{Schematic of the perpendicular laser cooling geometry. The black circle represents the approximate area occupied by 127 ions, which have a collective rotating motion at frequency $\omega_R$. Two parallel beams (not displayed here) are perpendicular to the X-Y plane. The
    perpendicular laser shown in red has a Gaussian intensity profile
    with a width of $W_y$, a displacement of $d$ from the trap center and it is directed along $x$ axis.}
  \label{fig:doppler_cooling}
\end{figure}

The in-plane motion is cooled by an additional laser directed along the
$x$ axis, i.e. perpendicular to the trap magnetic field. The arrangement of the planar cooling beam is
illustrated in Fig.~\ref{fig:doppler_cooling}. This
beam has a width $W_y$ and is displaced by a distance $d$ from
the center of the trap. To adjust for the rotational motion
of the ions at the planar beam location, the beam is detuned by
$\Delta_\perp=-\gamma_0/2+k\omega_Rd$ from the atomic transition.

The in-plane and out-of-plane degrees of freedom are only weakly
coupled to one another. Therefore, they are not typically in
thermal equilibrium. To study the temperature anisotropy, we
introduce parallel and perpendicular temperatures
\begin{equation}
    T_\parallel=\frac{m_{Be}}{Nk_B} \sum_{i =
    1}^N v_{z,i}^2  \, ,
\end{equation}
\begin{equation}
    T_\perp=\frac{m_{Be}}{2Nk_B} \sum_{i =
    1}^N \left ( v_{x,i}^2 + v_{y,i}^2 \right ) \,
\end{equation}

To simulate the Doppler cooling process, we initialize the ions in a finite
temperature state by setting their positions to the 2D crystal equilibrium,
with random velocities drawn from a
Maxwellian (Gaussian) distribution at a given starting temperature. We use a perpendicular cooling
beam with a peak saturation intensity of $S_{0}=1$ and a width of
$W_{y}=\SI{5}{\micro\meter}$ displaced by
$d=\SI{5}{\micro\meter}$ from the trap center. The axial cooling
beams have a uniform saturation intensity of $S_{0}=0.005$. We
deliberately choose a much smaller axial cooling beam intensity
because cooling of the out-of-plane motion is very efficient since all ions interact continuously with the parallel beams. Larger
axial cooling beam intensities would heat the in-plane degrees of
freedom due to recoil of the scattered photons, overpowering the planar cooling.

\begin{figure}
  \includegraphics[width=8cm]{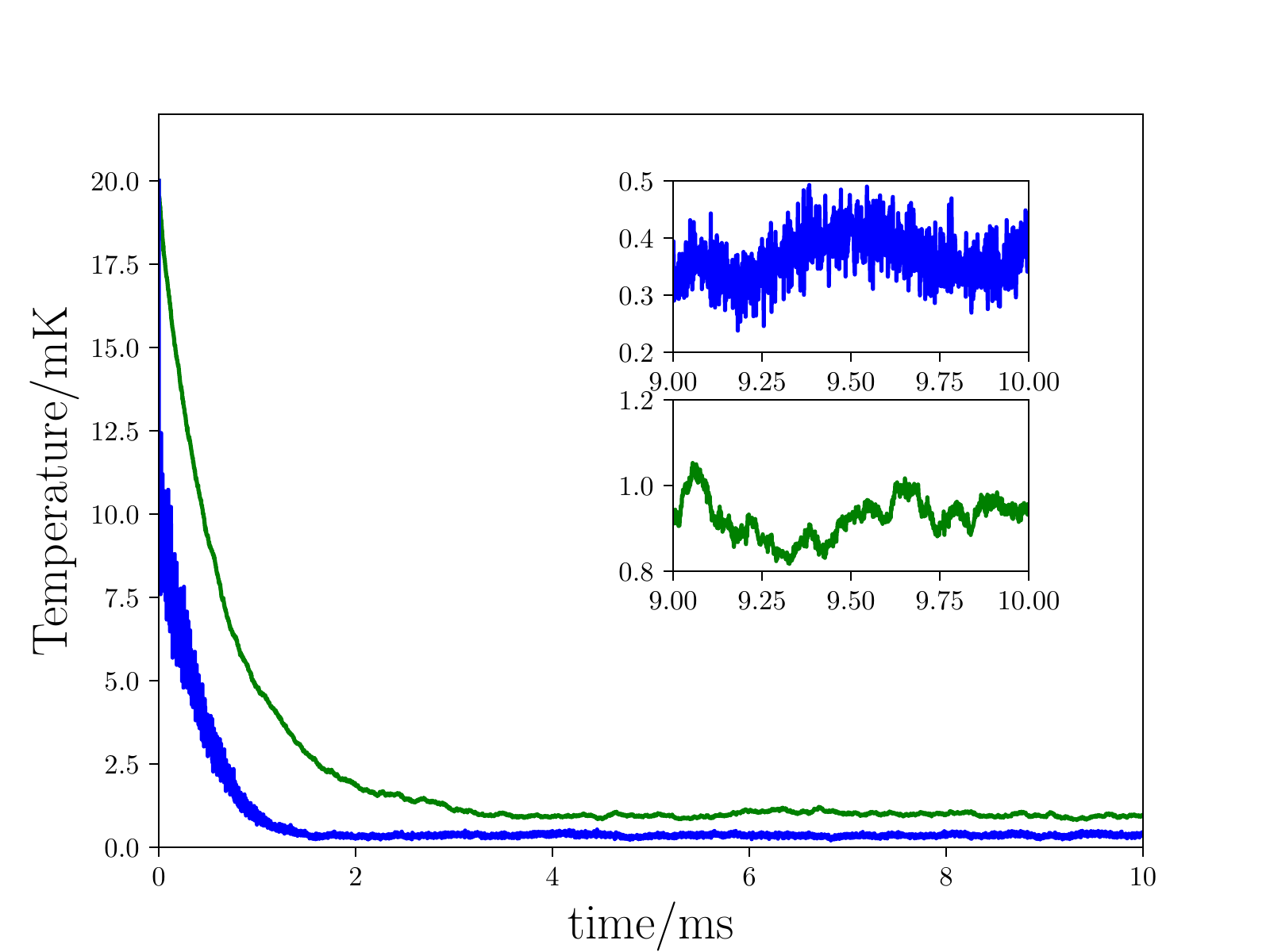}
  \caption{A simulation of Doppler cooling starting with an initial temperature $T=\SI{20}{\milli\kelvin}$. The blue line represents axial motion and the green line represents planar motion. The ion crystal achieves a steady-state temperature in qualitative agreement with experiment. The two inserts show a zoom-in of the axial and planar temperatures during the last millisecond of cooling. For these parameters (see text), the steady state axial temperature is $T_{\parallel}=\SI{0.37}{\milli\kelvin}$, and the planar temperature is $T_{\perp}=\SI{0.92}{\milli\kelvin}$.}
  \label{fig:cooling_sample}
\end{figure}

With all the cooling beams on, the system is left to evolve
for $\SI{10}{\milli\second}$. The time histories of the 
in-plane and out-of-plane temperatures during the simulation are
shown in Fig.~\ref{fig:cooling_sample}. For these simulation
parameters, both in-plane and out-of-plane temperatures relax to
a steady state on a time scale on the order of
$\SI{1}{\milli\second}$. 

\subsection{Measured steady-state axial and planar temperatures}

We now present steady-state temperature results. The axial temperature $T_\parallel$ can be well-diagnosed in the experiments~\cite{jordan2018}. Also, the theoretical Doppler cooling limit\cite{torrisi2016perpendicular,itano1988perpendicular} gives an estimate for $T_\parallel$ that agrees fairly well with temperatures achieved in the NIST experiment\cite{bohnet2016quantum,britton2012engineered,sawyer2014spin}.  The theoretical Doppler cooling limit for motion parallel to the magnetic field is
\begin{equation}
T_{\parallel,Doppler } = \frac{1}{3} \frac{\hbar\gamma_0}{k_{B}}=\SI{0.29}{\milli\kelvin}
\label{equ:cooling_limit}
\end{equation}
where $\gamma_0 =2\pi \cdot \SI{18}{\mega\hertz}$ is the atomic transition line-width for $^9\textrm{Be}^+$. This is the cooling limit predicted for laser cooling along one axis (in this case parallel to the magnetic field) with minimal recoil heating from the perpendicular laser beam~\cite{itano1982laser}.  This limit assumes isotropic scattering, which was also assumed in the numerical simulation of Doppler laser cooling.
We estimate the  steady-state  $T_{\parallel}$ by averaging the results with the lasers on continuously over the final ms, the time interval shown in the inserts of Fig. ~\ref{fig:cooling_sample}. We find steady state temperatures in Fig.~\ref{fig:cooling_sample} of $T_{\parallel}\approx\SI{0.37}{\milli\kelvin}$ and $T_{\perp}\approx\SI{0.92}{\milli\kelvin}$. The lowest $T_{\parallel}$ was obtained in a different simulation with $d=0$ (no planar beam offset) and was measured to be $\SI{0.28}{\milli\kelvin}\pm \SI{0.04}{\milli\kelvin}$, in good agreement with the Doppler cooling limit in Eq.~\ref{equ:cooling_limit}.
Also, it is important to note that the simulation shows fluctuations in the axial temperature to be approximately $\pm 15\%$.

We can use the simulation to measure the thermal energy in each axial mode and investigate if an equipartition assumption is valid.  Because the system is at very low temperature, nonlinear couplings are very weak.  Additionally, the simulations are for a finite time. Therefore, one might not expect the system to be in equipartition.
For an axial mode with frequency $\omega_m$, we can obtain the mode temperature $T_\parallel(\omega_m)$ from the spectrum
\begin{equation}
    T_{\parallel } (\omega_m) =
    \frac{m\omega_m^2\tau}{\pi}
    \int_{\rm res.} \hbox{d}\omega P_z(\omega)\;,
\end{equation}
where the integral is over a single resonance, and we evaluate the integral using the trapezoidal rule. Doppler cooling is turned off during the mode temperature recording time $\tau$.

%\begin{table}[width=8cm]
\begin{table}
  \begin{center}
    \begin{tabular}{|l|c|c|}
      \hline
      $m$ &$\omega_m$&$T_\parallel(\omega_m)/T_{\parallel, \rm Doppler}$\\
      \hline
      0&  \SI{1.580532}{\mega\hertz}&  \SI{1.05}{}\\
      1&  \SI{1.553344}{\mega\hertz}&  \SI{0.58}{}\\
      2&  \SI{1.552781}{\mega\hertz}&  \SI{0.31}{}\\
      3&  \SI{1.532620}{\mega\hertz}&  \SI{0.81}{}\\
      4&  \SI{1.532467}{\mega\hertz}&  \SI{0.81}{}\\
      5&  \SI{1.515915}{\mega\hertz}&  \SI{0.85}{}\\
      6&  \SI{1.515464}{\mega\hertz}&  \SI{0.94}{}\\
      7&  \SI{1.511197}{\mega\hertz}&  \SI{1.01}{}\\
    %  8&  \SI{1.501334}{\mega\hertz}&  \SI{2.621553E+01}{}&  \SI{4.373392E+00}{}\\
      \hline
    \end{tabular}
  \end{center}
  \caption{Mode temperatures of the first eight (starting with $m=0$) axial eigenmodes.}
  \label{tab:mt}
\end{table}

Table~\ref{tab:mt} shows the axial mode temperatures, relative to the theoretical parallel Doppler laser cooling limit, for the eight highest frequency axial modes for 127 ions. The axial mode temperatures shown in Table~\ref{tab:mt} are also obtained again with $d=0$ which gives the lowest parallel temperature steady state. The values are in the range of the Doppler cooling limit given the $\pm 15\%$ fluctuation, except for the two tilt modes (mode 1 and 2).  Higher mode numbers are harder to diagnose using Eq.~(26) due to the "grassy" spectrum in the range of $\SI{1.1}{\mega\hertz}$ to $\SI{1.5}{\mega\hertz}$ seen in Fig.~3.(b). The spectral peaks become very close to each other with increasing mode number making them difficult to differentiate. 
Additionally, the lowest $T_\perp$ is obtained by varying the $W_y$ and $d$ and found to be $T_{\perp}\approx\SI{1}{\milli\kelvin}$ with $W_y=d=\SI{5}{\micro\meter}$. Such small $W_y$ and $d$, compared to the radius of the crystal $R_c=\SI{70}{\micro\meter}$, allow the perpendicular beam to generate a modest torque which is straight forwardly balanced by the rotating wall\cite{torrisi2016perpendicular}. Additionally, the different $T_\perp$ and $T_\parallel$ demonstrate the temperature anisotropy and weak coupling between axial and planar directions.

\section{Summary}
\label{sec:conclusion}
We have described in detail a direct numerical simulation model of ultra-cold ions in a Penning trap with a rotating wall.  The simulation includes a Doppler cooling model based on the microscopic physics of resonance fluorescence with a realistic laser beam geometry.  Due to the low temperatures of interest, very good agreement is obtained with a linear eigenmodes analysis, in both the in-plane and out-of-plane directions.  The advantage of a direct numerical simulation is that weak nonlinear coupling between modes is accurately modeled, as well as the laser cooling/heating processes. Additionally, it is difficult to diagnose in-plane motion experimentally, so simulations can help in better interpreting and understanding this physics.

Such a model is very useful for understanding parametric dependences of the heating and cooling processes in the experiment, as well as the relevant time scales.   The Doppler cooling process was successfully incorporated in a molecular dynamics simulation for the first time, and the results were compared with existing theory.  Axial and planar temperatures were within the range of experimental results and more work is needed to make detailed comparisons with experiment.
 We varied the waist $W_y$ and offset $d$ of the  perpendicular laser beam to obtain the lowest planar temperature with the laser offset equal to the beam width  $d=W_y=\SI{5}{\micro \meter}\ll R_c\approx \SI{70}{\micro \meter} $. The fluid equilibrium model~\cite{torrisi2016perpendicular} predicts low in-plane temperatures for small $W_y$, with a very weak dependence on $d$ if $\Delta_\perp=-\gamma/2+k\omega_Rd$.  This prediction will be interesting to investigate, because increasing $d$ produces a large laser torque which, when balanced with a similar large torque from the rotating wall, produces shear and potential instabilities in the crystal.  
 %and  $T_\perp\left ( d, W_y \right )$ shows similar behavior as the cold fluid model\cite{torrisi2016perpendicular}. 

There are areas where the simulation model could be improved further. The Doppler cooling model treats photon-ion interactions instantaneously, neglecting the transition time.  We also neglect infrequent interactions with background neutrals and associated  ion loss, as well as, impurities and trap error fields.  These features of a more realistic experiment will be addressed in future work.

\section{Acknowledgment}
\label{sec:acknowledgment}

We thank Murray Holland and Athreya Shanker, JILA, Univ. of Colorado and Elena Jordan, NIST, for useful discussions.  Work supported in part by U.S. Department of Energy under award number DE-FG02-08ER54954.  This manuscript is a contribution of NIST and not subject to U.S. copyright. 

\bibliography{cite.bib}

\end{document}